\documentclass[aps,prl,twocolumn]{revtex4}
 
\begin{document}

\noindent
{\bf Comment on ``Current fluctuations in non-equilibrium diffusive
systems: an additivity principle''}\\

The main purpose of this comment is to point out that the results of
Ref.\ \cite{Derrida} hold without the need of either an
additivity principle, or scaling assumption which are foundational for
Ref.\ \cite{Derrida}.  We further comment on the range of validity of
the theory and the advantages of our approach \cite{uslong}.

In a recent Letter \cite{Derrida}, Bodineau and Derrida  find that the
``distribution of current fluctuations through a large one dimensional
system in contact with two reservoirs at unequal densities'' depends 
on only two parameters, the diffusion constant $D(\rho)$ and 
the noise density $\sigma(\rho)$, where $\rho$ is the charge density.
Their result is based on the assumptions of an ``additivity principle'' 
and a ``scaling hypothesis''.  The additivity principle postulates 
that a system may be decomposed into two independent subsystems, except 
that they are allowed to adjust the density $\rho$ at their contact 
in order to maximize the probability to observe a given integrated 
current. The scaling hypothesis postulates that the current cumulant 
generating function scales as $1/N$, where $N$ is the large system size. 

While the assumptions are interesting, it is natural to ask about
their microscopic origin, and in what situations the principles will
successfully describe the physics.  Therefore, it is important to have
an alternative approach which is based on microscopic conditions for
transport.  Such a theoretical approach has been formulated in Ref.\
\cite{uslong}, where the results of the paper \cite{Derrida} have been
derived independently with no assumption of the additivity principle
or of the scaling hypothesis.  Our method is related to an earlier
technique \cite{bertini}, although they did not consider transport
statistics and assume Gaussian fluctuations a priori.

The above additivity principle may be {\em derived} as the continuum
limit of our theory of transport statistics in an arbitrary classical
network, consisting of nodes storing charge and connectors carrying
statistically distributed currents \cite{usprl}.  Our approach
represents the current probability distribution function by a
stochastic path integral, whose validity depends on (i) a separation
of time scales requirement, prompting the Markov approximation, (ii)
the assumption of an effectively continuous charge, and (iii) the
saddle-point approximation.  The saddle-point approximation is well
controlled by a large parameter $\gamma$, the number of elementary
charges participating in the transport. The dominant contribution to
the generator of current cumulants, given by the saddle-point action
$S$, is proportional to $\gamma$, with corrections of order 1. Thus
the additivity principle may fail if $\gamma\sim 1$.  For exclusion
processes the assumption of effectively continuous charge is an
approximation, which may fail close to critical points.

Considering a lattice of $N\gg 1$ of nodes and taking a
continuum limit of the stochastic path integral, i.e. expanding the 
action $S$ to leading order in $1/N$, we have arrived \cite{uslong} 
at the following 1D diffusive field theory:
\begin{equation} S =
-\int_0^{\,t} \! dt'\! \int_0^L dz [{\lambda} {\dot
\rho} + D\rho'\lambda' -
(F/2)(\lambda')^2] ,
\label{fieldaction}
\end{equation}
where $\lambda$ is the ``counting field'', $D(\rho)$ is the diffusion
constant, and $F(\rho)$ is the noise density in our notation. In the
stationary limit, the solutions of the saddle-point equations of
(\ref{fieldaction}) with the boundary conditions $\lambda(0)=0$,
$\lambda(L)=\chi$ must be substituted back into the action to give the
generating function $S(\chi)$ of cumulants of transmitted charge.
This result is described in Eqs.\ (79-81) of \cite{uslong} and is
mathematically equivalent to the solution of Ref.\ \cite{Derrida}
giving the generator (7,8) as coupled, implicit integral equations,
after the said assumptions have been made.  For example, the field
equation (16) of Ref.\ \cite{Derrida} for the density $\rho$ is the
same as our equation (79) in Ref.\ \cite{uslong}.

We would like to stress that the additivity principle does not need to
be postulated. From the field-theoretic point of view, it is the
direct consequence of the space-time locality of the elementary
transport transitions (sources of noise), and of the existence of a
large parameter, the number of particles participating in transport.
The fact that the transport statistics is characterized by only two
parameters $D(\rho)$ and $F(\rho)$ originates from the central limit
theorem, and is a consequence of the large $N$ limit which removes 
high powers of $\lambda$ from the action and leads to the Gaussian 
character of the local noise.

The advantage of the field-theoretic approach is not only that it
shows the limits of the applicability of the theory, but also that it
provides a framework for the perturbation expansion of current
cumulants \cite{uslong}, investigation of time-dependent phenomena
\cite{sp1}, and treatment of multiple dimensions \cite{uslong}.

\vspace{.5cm}
\noindent
E.\ V.\ Sukhorukov and A.\ N.\ Jordan,\\
D\'epartement de Physique Th\'eorique\\ Universit\'e de Gen\`eve\\
CH-1211 Gen\`eve 4, Switzerland\\
\\
Date: June 15, 2004

\end{document}